\documentclass{aa}  

\usepackage{graphicx}
\usepackage{txfonts}
\usepackage{verbatim}
\usepackage{natbib}
\usepackage{array}
\usepackage{threeparttable}
\usepackage{xcolor}
\usepackage{multirow}

\def\msun{M$_{\odot}$}
\def\msol{M$_{\odot}$}

\begin{document}

    \title{Constraining the overcontact phase in massive binary evolution - II. Period stability of known O+O overcontact systems}
    \titlerunning{Period stability of O+O overcontacts}
    \authorrunning{M. Abdul-Masih et al.}

    \author{Michael Abdul-Masih \inst{1} \and
            Ana Escorza \inst{1} \and
            Athira Menon \inst{2,3,4} \and
            Laurent Mahy \inst{5} \and
            Pablo Marchant \inst{6}
          }

    \institute{European Southern Observatory, Alonso de Cordova 3107, Vitacura, Casilla 19001, Santiago de Chile, Chile\\
    \email{michael.abdul-masih@eso.org}
    \and
    Instituto de Astrofísica de Canarias, 38 200 La Laguna, Tenerife, Spain 
    \and
    Universidad de La Laguna, Universidad de La Laguna, 38 205 La Laguna, Tenerife, Spain
     \and
    Argelander-Institut für Astronomie, Universität Bonn, Auf dem Hügel 71, 53121 Bonn, Germany
    \and
    Royal Observatory of Belgium, Avenue Circulaire/ Ringlaan 3, B-1180 Brussels, Belgium
    \and
    Institute of Astronomy, KU Leuven, Celestijnenlaan 200 D, 3001 Leuven, Belgium
    }

   \date{}

 
  \abstract
   {Given that mergers are often invoked to explain many exotic phenomena in massive star evolution, understanding the evolutionary phase directly preceding a merger, the overcontact phase, is of crucial importance.  Despite its importance, large uncertainties exist in our understanding of the evolution of massive overcontact binaries.}
   {We aim to provide robust observational constraints on the future dynamical evolution of massive overcontact systems by measuring the rate at which the periods change for a sample of six such objects. Furthermore, we aim to investigate whether the periods of unequal mass systems show higher rates of change than their equal mass counterparts as theoretical models predict.}
   {Using archival photometric data from various ground- and space-based missions covering up to $\sim$40 years, we measure the periods of each system over several smaller time spans.  We then fit a linear regression through the measured periods to determine the rate at which the period is changing over the entire data set.}
   {We find that all of the stars in our sample have very small period changes and that there does not seem to be a correlation with the mass ratio.  This implies that the orbital periods for these systems are stable on the nuclear timescale, and that the unequal mass systems may not equalize as expected.}
   {When comparing our results with population synthesis distributions, we find large discrepancies between the expected mass ratios and period stabilities.  We find that these discrepancies can be mitigated to a degree by removing systems with shorter initial periods, suggesting that the observed sample of overcontact systems may originate from binary systems with longer initial orbital periods.}

   \keywords{binaries: close --
                Stars: massive --
                Stars: evolution --
                Techniques: photometric
               }

   \maketitle
%

\section{Introduction}

    With a binary fraction of $\sim$ 100\%, the presence of a companion plays a crucial role in the evolution of massive stars \citep[e.g., ][]{Sana2011,Duchene2013, Moe2017}.  Throughout their lives, it is expected that approximately 70\% of massive stars will interact with a companion \citep[e.g., ][]{Sana2012} and of which 40\%  (24\% of all massive stars) will evolve through an overcontact phase \citep[e.g., ][]{Pols1994, Wellstein2001, deMink2007}. Despite this, very few massive overcontact systems are known \citep[see e.g., ][]{Leung1978, Popper1978, Hilditch2005, Penny2008, Lorenzo2014, Lorenzo2017, Almeida2015, Martins2017, Mahy2020a, Janssens2021}.
    
    Despite the rarity of these systems, the overcontact phase can be of crucial importance in the evolution of massive binary systems. The unique geometry and strong binary interactions during this phase make the internal processes difficult to accurately constrain \citep[see e.g., ][]{Fabry2022}. Depending on the treatment of these internal processes and the rate of mass transfer as a binary system first comes into contact, systems evolving through this phase can have drastically different end products.  For example, objects such as magnetic massive stars \citep{Schneider2019}, Be stars \citep{Shao2014}, Luminous Blue Variables \citep{Justham2014, Smith2018}, blue stragglers \citep{Eggen1989,Mateo1990} and peculiar Type-II supernovae like SN-1987A \citep{Podsiadlowski1992, Menon2017, Urushibata2018} have all been postulated to be the direct result of massive binary mergers.  Alternatively, if the conditions are right (i.e. efficient internal mixing), theoretical studies predict that overcontact systems may be able to avoid merging while on the main sequence, instead forming double black hole binary systems and eventually gravitational wave sources via the chemically homogeneous evolution pathway \citep{deMink2016, Mandel2016, Marchant2016, duBuisson2020, Riley2021}.
    
    An important question when considering the future evolution of a massive overcontact binary system is whether it is evolving on a nuclear timescale, implying that the system is relatively stable, or on a thermal timescale, implying that the system is unstable and will most likely either merge or separate \citep{Pols1994}. Due to their extremely short-lived nature, observing a thermal-timescale overcontact system is expected to be very unlikely, so it is often assumed that the known massive overcontact systems are evolving on the nuclear timescale. Theoretical studies focused on stable massive overcontact binaries indicate that these systems should very quickly equalize in mass, and then continue to evolve on a nuclear timescale \citep{Marchant2016, Menon2021}. Observationally, however, most of the known massive overcontact binaries are found in unequal mass systems.
    
    The discrepancy between the observed and expected mass ratios in combination with the lower than expected number of known massive overcontact binaries when compared with predictions from population synthesis studies \citep[e.g.][]{Langer2020, Menon2021} lead to several interesting open questions.  Is the contact phase less stable and therefore shorter lived than we expect?  Are we preferentially observing systems before they equalize in mass or is our prediction that these systems equalize flawed? By investigating how the period changes over several years, we can begin to answer some of these questions.

    In this paper, we combine archival photometric data sets obtained over a long period of time to investigate the period stability of known massive overcontact systems. By determining how quickly the orbital period is changing, we can determine whether these systems are evolving via the nuclear or the thermal time scale.  Further, we can determine whether these systems are in the process of equalizing in mass or if they are evolving as long-lived but unequal mass overcontact binaries.  In Sect. \ref{Sample}, we discuss our sample selection and the available photometry for each source as well as our data reduction techniques (when applicable). In Sect. \ref{Methods} we detail our period determination procedure and how we calculate the period stability.  We present our results in Sect. \ref{Results}, and we discuss the implications of our findings in Sect. \ref{Discussion}.  Finally, Sect. \ref{Conclusions} summarizes our findings and discusses future prospects.


\section{Sample and Observations}\label{Sample}

    \begin{figure*}[ht]
    \centering
    \includegraphics[width=\linewidth]{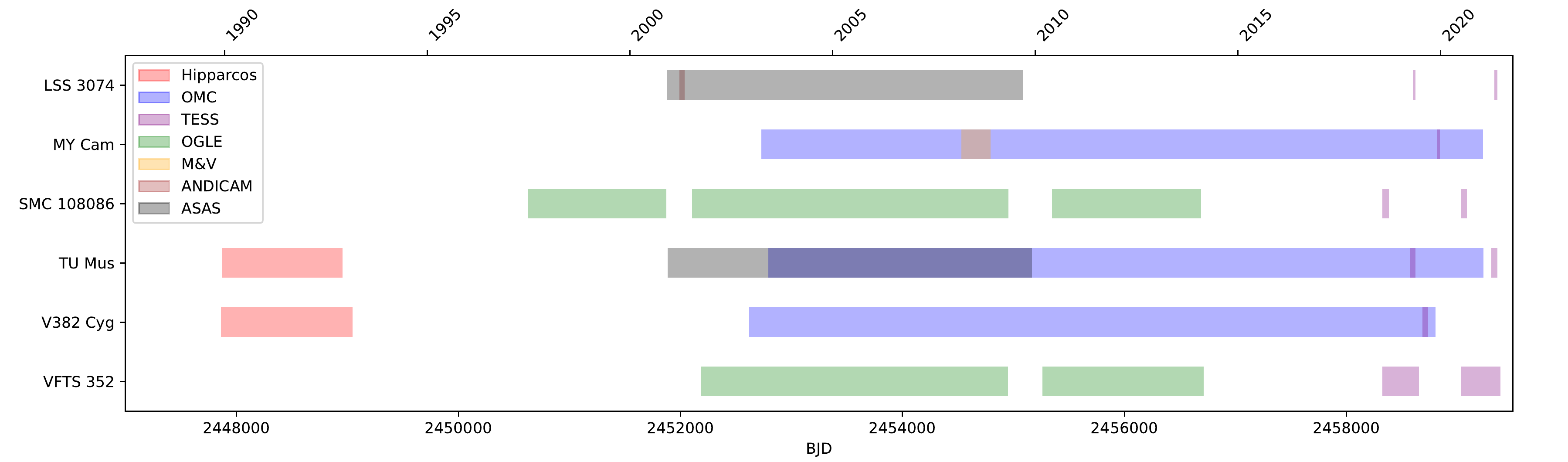}
    \caption{Overview of the photometric data available for each object in our sample.  Each colored shaded region corresponds to a different instrument or data set.  Red indicates Hipparcos, blue indicates INTEGRAL OMC, green indicates OGLE, black indicates ASAS, brown indicates ANDICAM, yellow indicates data from \citet{Lorenzo2014} and finally, data from TESS are indicated in purple.  Note that these ranges indicate the date ranges of the respective data sets, some of which are sporadic and without a regular cadence.}
    \label{fig:data_overview}
    \end{figure*}

    \begin{table*}
    \caption{Selected orbital parameters for each of the systems in our sample.  The last column indicates the reference from which the orbital solutions were derived.   Note that while the other parameters come directly from each of these papers, the fillout factors were calculated using Eq. \ref{eq:fillout_factor} as described in Sect. \ref{Sample}.}
    \centering 
    \begin{tabular}{ccccccc}
    \hline\hline
     & $P_\mathrm{orb}$  & $M_1$ & $M_2$ & $q$ ($M_1/M_2$) & $f$ & reference \\
     & [d] & [\msol] & [\msol] & & & \\
    \hline  
    LSS 3074 & 2.1852 & 17.2 $\pm$ 1.4 & 14.8 $\pm$ 1.1 & 0.86 $\pm$ 0.04 & 1.05 & \citet{Raucq2017}\\
    MY Cam & 1.1754514 & 37.7 $\pm$ 1.6 & 31.6 $\pm$ 1.4 & 0.84 $\pm$ 0.03 & 1.01 & \citet{Lorenzo2014}\\
    SMC 108086 & 0.8830987 & 16.9 $\pm$ 1.2 & 14.3 $\pm$ 1.7 & 0.85 $\pm$ 0.06 & 1.70 & \citet{Hilditch2005}\\
    TU Mus & 1.387282 & 16.7 $\pm$ 0.4 & 10.4 $\pm$ 0.4 & 0.623 $\pm$ 0.009 & 1.12 & \citet{Penny2008}\\
    V382 Cyg & 1.885545 & 26.1 $\pm$ 0.4 & 19.0 $\pm$ 0.3 & 0.727 $\pm$ 0.005 & 1.10 & \citet{Martins2017}\\
    VFTS 352 & 1.1241452 & 28.9 $\pm$ 0.3 & 28.6 $\pm$ 0.3 & 0.99 $\pm$ 0.10 & 1.28 & \citet{Almeida2015}\\
    \hline
    \end{tabular}
    \label{table:orb_sol}
    \end{table*}

    Since the goal of this investigation is to characterize the period change in massive overcontact systems, we select our sample based on a set of criteria designed to ensure that we remove as many biases as possible.  These are detailed below:
    
    \begin{itemize}
    \item The optimal solution for the system must be an overcontact configuration and further, this must have been determined via a combined photometric and radial velocity fit. Ensuring that the system is in an overcontact configuration is of the utmost importance since semidetached and detached systems with ellipsoidal deformations will have different period evolutions and will thus probe different physical effects than those that dominate during the overcontact phase.
    \item The system must not be in a confirmed triple or higher order multiple system unless we can ensure that the additional components are far enough from the binary such that they have a negligible effect on the dynamics of the system \citep[see e.g., ][]{Toonen2016}.  The presence of a nearby third object ($P_\mathrm{out} \lesssim 10$ yr for massive overcontact systems) is known to alter the period and orbital parameters of the inner binary system via von Zeipel-Kozai-Lidov oscillations \citep[vZKL; ][]{vonZeipel1910, Kozai1962, Lidov1962}.  These perturbations could bias the period variation measurements and for this reason, these systems are excluded from our sample.
    \item If the photometric data of the system is contaminated with other periodic signals, the signature of the binary must be the dominant signal.  
    \item Both of the system's components must be main sequence O-type stars.  This criterion is meant to ensure that the sample is as complete as possible in the given spectral range, while also limiting the sample size to a manageable amount.
    \end{itemize}
     
    With the above criteria, our final sample consists of 6 objects that are spread over different metallicity regimes including the Milky Way, Large and Small Magellanic Clouds.  These systems and their available photometric data are discussed in detail below.  An overview of the photometric data used for each target as well as the time bases that the different datasets cover is presented in Fig. \ref{fig:data_overview}.  Additionally, the most relevant parameters to this study including the periods, mass ratios and fillout factors are summarized in Table \ref{table:orb_sol}.
    
    For the purposes of this study, we define the primary as the currently more massive component and the mass ratio ($q$) as the mass of the secondary over the mass of the primary such that $q \leq 1$.  The fillout factor, which is a measure of the degree to which a system is overfilling its Roche lobes, has several different definitions in the literature.  Here, we adopt the definition of the fillout factor $f$ from \citet{Mochnacki1972}, which states:
  	\begin{equation}\label{eq:fillout_factor}
      f = \frac{\Omega_{n,1} - \Omega_n}{\Omega_{n,1} - \Omega_{n,2}} + 1,
  	\end{equation}
    where $\Omega_{n,1}$ and $\Omega_{n,2}$ denote the potential of the surface passing through L1 and L2 respectively, and $\Omega_n$ indicates the measured surface potential of the system.  In this definition, an overcontact system has a fillout factor $1 < f < 2$, with higher fillout factors corresponding to systems in deeper contact.  Since the degree of deformation for the systems in our sample are not presented in a consistent way throughout the literature, we compute the fillout factor according to the above definition for each object in our sample to ensure homogeneity.

\subsection{LSS 3074}

LSS 3074 was initially characterized as a contact system by \citet{Raucq2017} and is located in the Milky Way.  With a fillout factor of 1.05 the system is just barely in contact, however, the photometric analysis strongly favors a contact configuration over a semidetached configuration.  The period was measured to be 2.1852 days, making it the longest period system in our sample. This, in combination with its masses of 17.2 and 14.8 \msol, imply that it may be slightly more evolved than the rest of our sample.  While the spectral types of both components appear to be solidly in the O-type regime, the anomalous combination of certain spectral features did not allow \citet{Raucq2017} to firmly determine a spectral type for each component. 

The photometric data set for LSS 3074 consists of data from the All Sky Automated Survey (ASAS), data from A Novel Dual Imaging Camera (ANDICAM) and two sectors of data from the Transiting Exoplanet Survey Satellite (TESS).  The data from ANDICAM were collected between March and May of 2001 and were observed in the Johnson B-, V-, R- and I-bands \citep{Raucq2017}.  The ASAS data were collected sporadically over a $\sim$ 9 year period between 2000 and 2009 and were observed in the V-band \citep{Pojmanski1997, Pojmanski2002, Pojmanski2003, Pojmanski2004, Pojmanski2005b, Pojmanski2005a}.  Since LSS 3074 is a southern object, it was observed during the first and third year of TESS mission with data in sectors 11 and 38, respectively \citep{Ricker2015}. It should be noted that there is also data from the International Gamma-Ray Astrophysics Laboratory Optical Monitoring Camera (INTEGRAL-OMC) available for the target, however, the quality of the light curve was not good enough to allow us to detect a statistically significant peak near the orbital frequency so we do not include it in this analysis. 

\subsection{MY Cam}
MY Cam is located in the Milky Way and was first characterized as a contact system by \citet{Lorenzo2014}.  With component masses of 37.7 and 31.6 and spectral types of O5.5 and O7 respectively, it is the most massive overcontact system currently known.  Its period was measured to be $\sim$ 1.175 days and it has a mass ratio of 0.84.  Of all of the systems in our sample, MY Cam has the lowest measured fillout factor at only 1.01, meaning that it just barely qualifies as an overcontact system.  

The photometric data set for MY Cam consists of data from INTEGRAL-OMC and TESS as well as data from two private telescopes. The INTEGRAL-OMC data were observed in the Johnson V-band and were collected sporadically over an $\sim$ 18 year time frame between 2003 and 2021 \citep{Alfonso-Garzon2012}.  Unfortunately, only one sector of TESS data is available, which was observed in sector 19 during the second year of the TESS mission.  In addition to these, photometric data were collected from two private telescopes during a 6 month period in 2008.  These two telescopes were a Meade LX200 and a Vixen VISAC and observed in the Johnson R-band \citep{Lorenzo2014}.  Since the telescope and instrument names were not provided, we refer to this dataset as M\&V henceforth reflecting the telescope models from which the data were collected.

\subsection{OGLE SMC-SC10 108086}
OGLE SMC-SC10 108086 (SMC 108086 henceforth) was first characterized as a contact system by \citet{Hilditch2005}, and as its name suggests, it is located in the Small Magellanic Cloud (SMC).  The primary and secondary components have spectral types of O9 and O9.5 respectively and their locations on the Hertzsprung-Russell diagram (HRD) indicate that they are very close to the Zero-Age Main Sequence \citep{Abdul-Masih2021}.  With a fillout factor of 1.7 and a period of around 0.88 days \citep{Pawlak2016}, it is both the deepest massive overcontact system currently known and the shortest period system in our sample.  This, in combination with its mass ratio of 0.85, makes it an ideal test case for this investigation.

The photometric dataset for SMC 108086 consists of both Optical Gravitational Lensing Experiment (OGLE) and TESS data.  As part of the OGLE II, III and IV campaigns, it was observed sporadically over a total time span of $\sim$ 16 years \citep{Udalski1997, Udalski2008, Udalski2015, Szymanski2005}.  While only I-band data is available for OGLE II, it was observed in both the I- and V-bands during OGLE III and IV.  Being in the southern hemisphere, it was observed during the first and third year of TESS with a total of 4 sectors of data available (sectors 1, 2, 27, and 28).  

\subsection{TU Mus}
Along with V382 Cyg, TU Mus was one of the first massive overcontact systems identified and is located in the Milky Way.  It was originally characterized as a contact system by \citet{Andersen1975}, and has been studied extensively since then \citep[e.g., ][]{Stickland1995, Terrell2003, Linder2007, Qian2007, Penny2008}.  It has a period of around 1.387 days and a fillout factor of 1.12, and with a mass ratio of 0.623, it is the most unequal mass system in our sample. While it is universally agreed upon that the primary is an O-type star, there is some ambiguity in the literature as to the status of the secondary; some sources claim that it is a late O-type star \citep[e.g., ][]{Terrell2003, Penny2008} while others claim that it's spectral type is early B \citep[e.g., ][]{Sota2014, MaizApellaniz2016}. It is also important to note that \citet{Qian2007} found evidence of a third object gravitationally bound to the system, but given its long period ($\sim$47 years) and low component mass, it is expected to have a negligible effect on the dynamics of the inner contact system.  Based on the parameters of the system, the vZKL oscillations are expected to operate on timescales of $\sim$0.9 Myr \citep[see Eq. 24 in][]{Toonen2016}

The photometric dataset for TU Mus consists of data from Hipparcos, ASAS, INTEGRAL-OMC and TESS.  The Hipparcos data \citep{Perryman1997} were collected between December 1989 and November 1992, and were observed in the Hipparcos passband ($Hp$).  The ASAS data were collected sporadically over a $\sim$ 9 year period between December 2000 and December 2009 and were observed in the V-band.  The INTEGRAL-OMC data were collected in the Johnson V-band between 2003 and 2021. Finally, there are four sectors of TESS data available, two sectors (11 and 12) in the first year and two sectors (37 and 38) in the third year of the TESS campaign.  

\subsection{V382 Cyg}
V382 Cyg was first identified and characterized in the late 1970's \citep{Cester1978, Popper1978} and has been the subject of numerous studies since then \citep[e.g., ][]{Popper1991, Harries1997, Burkholder1997, Degirmenci1999, Qian2007, Yasarsoy2013}.  Located in the Milky Way, the primary and secondary components have spectral types of O6.5 and O6 respectively.  Recently, \citet{Martins2017} reanalyzed the system and updated the orbital parameters, reporting an orbital period of $\sim$ 1.89 days with a fillout factor of 1.10 and a mass ratio of 0.727.  Despite its low fillout factor, recent spectroscopic observations of this system indicate potentially high levels of mixing between the two components, giving further evidence that the system is indeed in a contact configuration \citep{Abdul-Masih2021}.  As with TU Mus, \citet{Qian2007} found evidence that V382 Cyg has a tertiary component, but its period and mass suggest that it is likely to have a negligible effect on the dynamics of the contact system with a vZKL oscillation timescale of $\sim$ 0.8 Myrs.  This was later confirmed by \citet{Yasarsoy2013}, who updated the period to be $\sim$ 43 years. 

The photometric data set for V382 Cyg is comprised of data from Hipparcos, INTEGRAL-OMC and TESS.  The data from the Hipparcos Catalog were observed between October 1989 and February 1993 and were observed in the Hipparcos passband.  The data from the INTEGRAL-OMC on the other hand were observed sporadically over a $\sim$18 year time frame between 2002 and 2019 and were observed in the Johnson V-band.  In addition to these, V382 Cyg was observed by TESS in sectors 14 and 15 during the second year of the TESS mission.


\subsection{VFTS 352}
VFTS 352 is located in the Large Magellanic Cloud (LMC henceforth) and was first characterized by \citet{Almeida2015}.  The nearly twin components ($q=0.99$) have masses of $\sim$29 \msun\ and have spectral types of O4.5 and O5.5, making it the earliest overcontact system currently known \citep{Walborn2014, Almeida2015, Almeida2017, Mahy2020a, Mahy2020b}.  Its high component masses, short period ($\sim$1.124 days) and relatively high fillout factor (1.28) make it a promising candidate for a gravitational wave progenitor \citep{deMink2016, Mandel2016, Marchant2016, Abdul-Masih2019, Abdul-Masih2020b, Abdul-Masih2021}. 

The photometric data set for VFTS 352 is comprised of data from the OGLE-III and -IV campaigns as well as TESS.  The data from OGLE III and IV were collected sporadically over a total time span of $\sim$ 13 years between 2001 and 2014 in both the I- and V-bands. Given its location in the LMC, VFTS 352 fell in TESS's continuous viewing zone meaning that it was observed for the entirety of the first and third years.

\subsection{Rejected systems}
Several other O-type overcontact systems are known, but these were not included in our sample for various reasons.  LY Aur is a known triple with vZKL oscillation timescales on order of $\sim$ 0.2Myrs \citep{Stickland1994, Zhao2014}.  Given the short oscillation timescale, we reject it from our sample.  
V729 Cyg is long period ($\sim$ 6.6 days), evolved overcontact system that is no longer on the main sequence so it is not included \citep{Antokhina2016}.
OGLE-SMC-ELC-4690 is thought to be a contact system, but no combined photometric and radial velosity fit has been performed on the object.  Furthermore, it has a known triple companion on a relatively close orbit \citep{Zasche2017}.
BAT 99-126 is a higher order system that contains a O-type contact system, however the orbital configuration of the system is not known so it is rejected \citep{Janssens2021}.
HD 64315 is a quadruple system containing two pairs of close binaries, one of which is in a contact configuration.  Unfortunately, the separation between the two pairs of binaries is not known so we do not include it in our sample \citep{Lorenzo2017}.
Finally, UW CMa appears to be a contact system, but the light curve has some unexplained features in it which makes the fitting unreliable.  So far, no reliable orbital solution has been found \citep{Leung1978,Antokhina2011}.

\subsection{Photometric data preparation}
While most of the photometric data used in this investigation were already reduced, some needed to be cleaned.  Specifically, in the case where quality flags were provided, we removed all data points that had bad quality flags following the individual recommendations of each data set.  For the data sets without quality flags, we removed obvious outliers.

\begin{figure}[t]
    \centering
    \includegraphics[width=\linewidth]{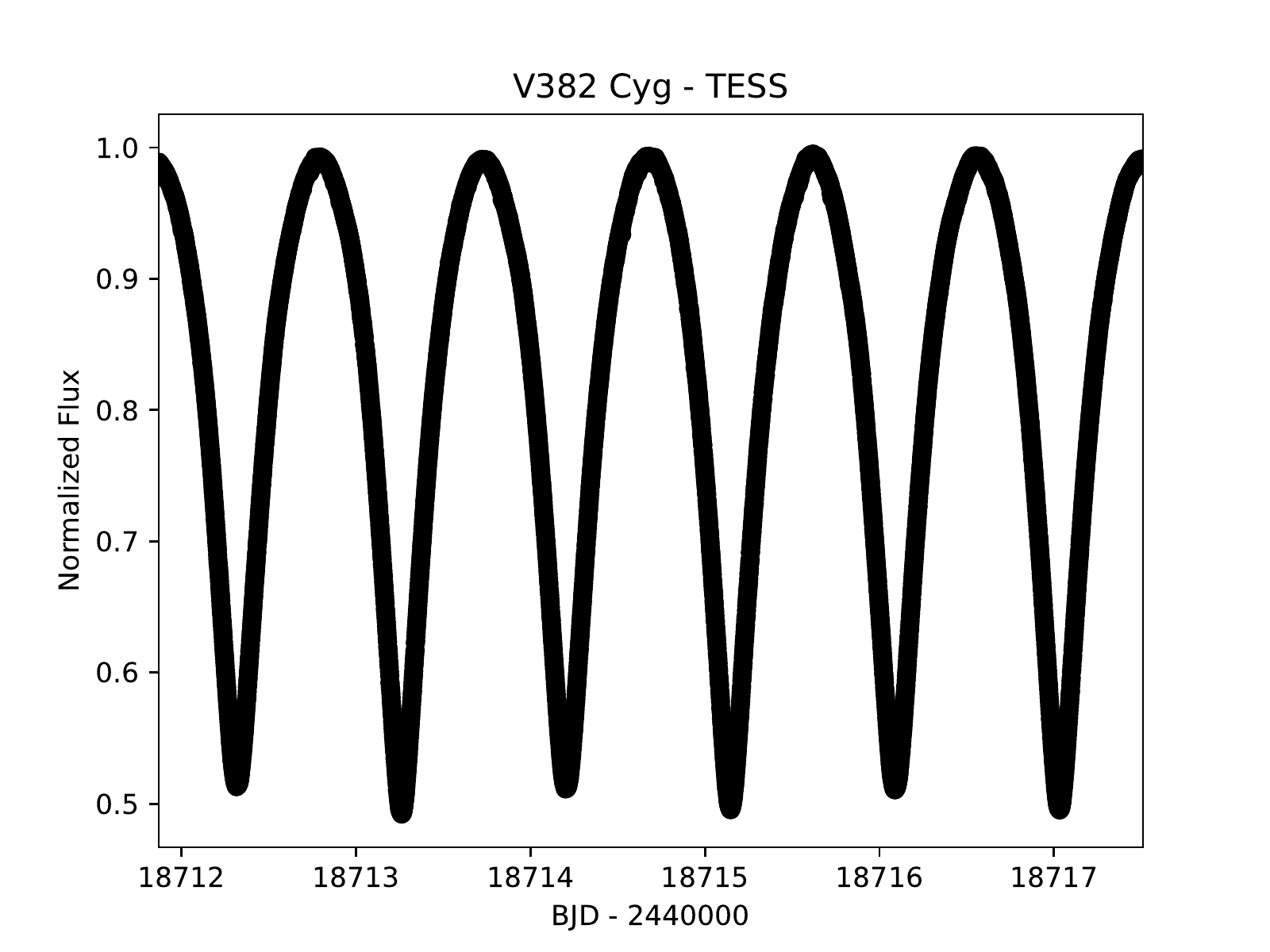}
    \caption{Portion of the TESS light curve associated with V382 Cyg}
    \label{fig:LC}
    \end{figure}

In the case of TESS, only some of the objects in our sample had reduced light curves associated.  While TESS is a nearly all sky survey, only some of the many stars observed have been reduced with the official TESS pipeline \citep[SPOC; ][]{Jenkins2016}.  Of the six stars in our sample, only V382 Cyg and TU Mus have SPOC light curves, so for these objects we use the available light curves (see Fig. \ref{fig:LC} as an example of the TESS light curve for V382 Cyg). For the four remaining sources, we utilize \textsc{lightkurve} \citep{LightkurveCollaboration2018} to aid in the extraction.  

\textsc{lightkurve} is a Python package designed for the retrieval and extraction of Kepler, K2 and TESS light curves.  From the full frame image, we first created a 9 x 9 pixel cutout centered on the source in question \citep{Brasseur2019}. We then created a mask, which only includes the central pixel of the 9 x 9 cutout and generate a light curve from this mask.  We choose to use only the central pixel in all cases to remain consistent between objects and sectors and to minimize the chances of contamination. VFTS 352 and SMC 108086 are located in crowded fields and given the TESS pixel size, eliminating contamination entirely is not possible.  Since no additional periodicities were found in the light curves of these objects, and since we are only concerned with the period, the presence of third light in these objects is not problematic for our specific science case. Once the light curves were extracted, we removed NaNs and outliers, and we detrended the resulting light curve using the \textsc{lightkurve} flatten function.   In some cases, there were trends at the beginning or end of the sectors as well as just before and after the mid sector downlinks.  In these cases, we remove the spurious points.


    \begin{figure}[t]
    \centering
    \includegraphics[width=\linewidth]{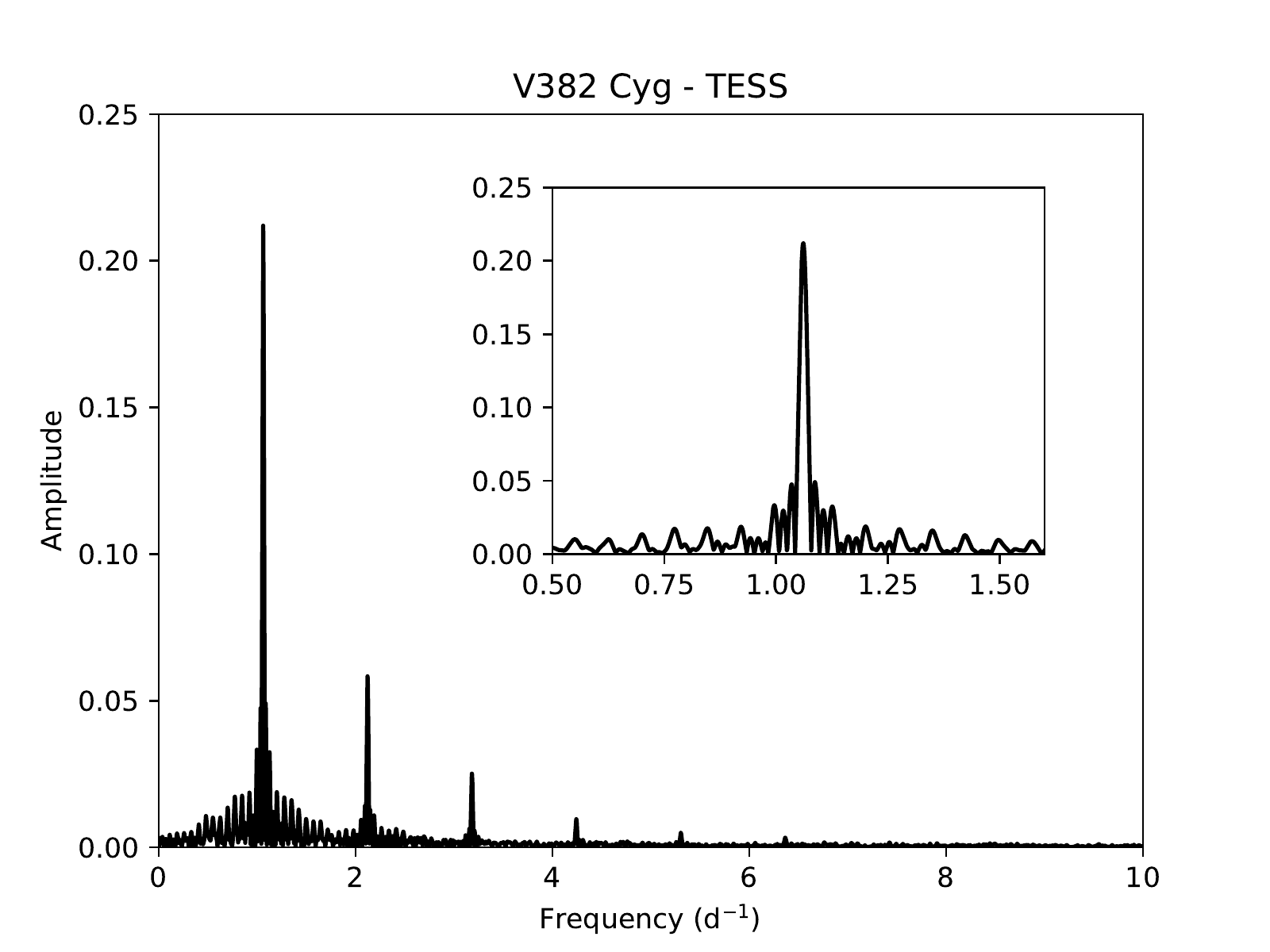}
    \caption{Fourier spectrum of the TESS light curve associated with V382 Cyg.  The inset shows a zoom in around the dominant frequency. The other peaks visible in the periodogram corresponds to harmonics of the fundamental frequency. }
    \label{fig:FT}
    \end{figure}

\section{Methods}\label{Methods}

\subsection{Period determination with \textsc{PERIOD04}}
In order to accurately determine the orbital period ($P_\textrm{orb}$) from each photometric data set, we used the software package \textsc{PERIOD04} \citep{Lenz2005}. This tool, based on classical Fourier Analysis Techniques, is especially dedicated to the statistical analysis of large astronomical data sets containing gaps. Using \textsc{PERIOD04}, we computed the frequency spectrum of each light curve (see Fig. \ref{fig:FT} as an example of the fourier spectrum of the TESS light curve for V382 Cyg) and identified the dominant periodicity in each data set. Since the orbital periods of our targets are known (see Table \ref{table:orb_sol}), we could easily determine if the dominant frequencies detected by the software were the true periods, a fraction/multiple of this, or a different periodicity present in the data. When the dominant frequency was not associated to the known period, we pre-whitened its contribution from the original data and continued extracting and pre-whitening frequencies until we could measure the period.

The uncertainties associated with each measured frequency were calculated by means of Monte Carlo simulations computed with \textsc{PERIOD04}. For each data set, we generated 1000 simulated time series with the times of observation matching those from the real data and the magnitudes (or intensities) were calculated from the magnitudes predicted by the best fit plus Gaussian noise. For every time string, a least-squares calculation was performed and the frequency uncertainty was calculated from the distribution of the Monte Carlo results.

The frequency peaks associated to the orbital periods were easily identified in all data sets with signal-to-noise ratios ranging between $\sim$5, for the shortest and most sparsely covered data sets, and $\sim$40, for the frequency spectrum of the TESS light curves. For all data sets and all targets, the orbital frequencies that we found corresponded to double the true orbital frequencies, which is expected given the symmetry shown by overcontact binary light curves (see Fig. \ref{fig:LC}).

Since our data sets vary widely in both cadence and time base, we treat some data sets slightly differently than others.  While the period determination process used is the same, some are split into smaller data subsets in order to avoid period smearing.  For example, due to the long time base ($\sim$20 years) and sporadic nature of the observations, we split the INTEGRAL-OMC data sets in half and analyze each independently.  Similarly, we treat data from OGLE II, III and IV separately and determine independent periods for each.  Finally, due to the biennial nature of the TESS mission, the TESS data set is divided by year.

\subsection{Determination of the change in period ($\dot{P}$)}
Once the periods associated with each dataset were determined, we fit a linear regression through the data for each object to determine the overall change in period ($\dot{P}$).  In some cases, multiple filters or apertures were observed simultaneously for a given data set (e.g. ASAS and ANDICAM), so to avoid unfairly weighing these data sets, we only include the aperture or filter that returned the lowest sigma from the period determination step in the linear fit.  In the case of OGLE, since the I- and V-band observations were not taken simultaneously we include both as distinct data sets when available.

To avoid correlations between the two free parameters in the linear regression (namely the slope and the y-intercept), we offset the times such that a time of 0 corresponds to the midpoint between the first and last central Barycentric Julian Date (BJD) for each object.  Here we define the central BJD of each data set as the midpoint of the observations. We optimize the two free parameters using the 'curve\_fit' function of the \textsc{SciPy} package, which utilizes non-linear least squares to fit \citep{Virtanen2020}.  

\section{Results}\label{Results}

\begin{figure}[t]
    \centering
    \includegraphics[width=\linewidth]{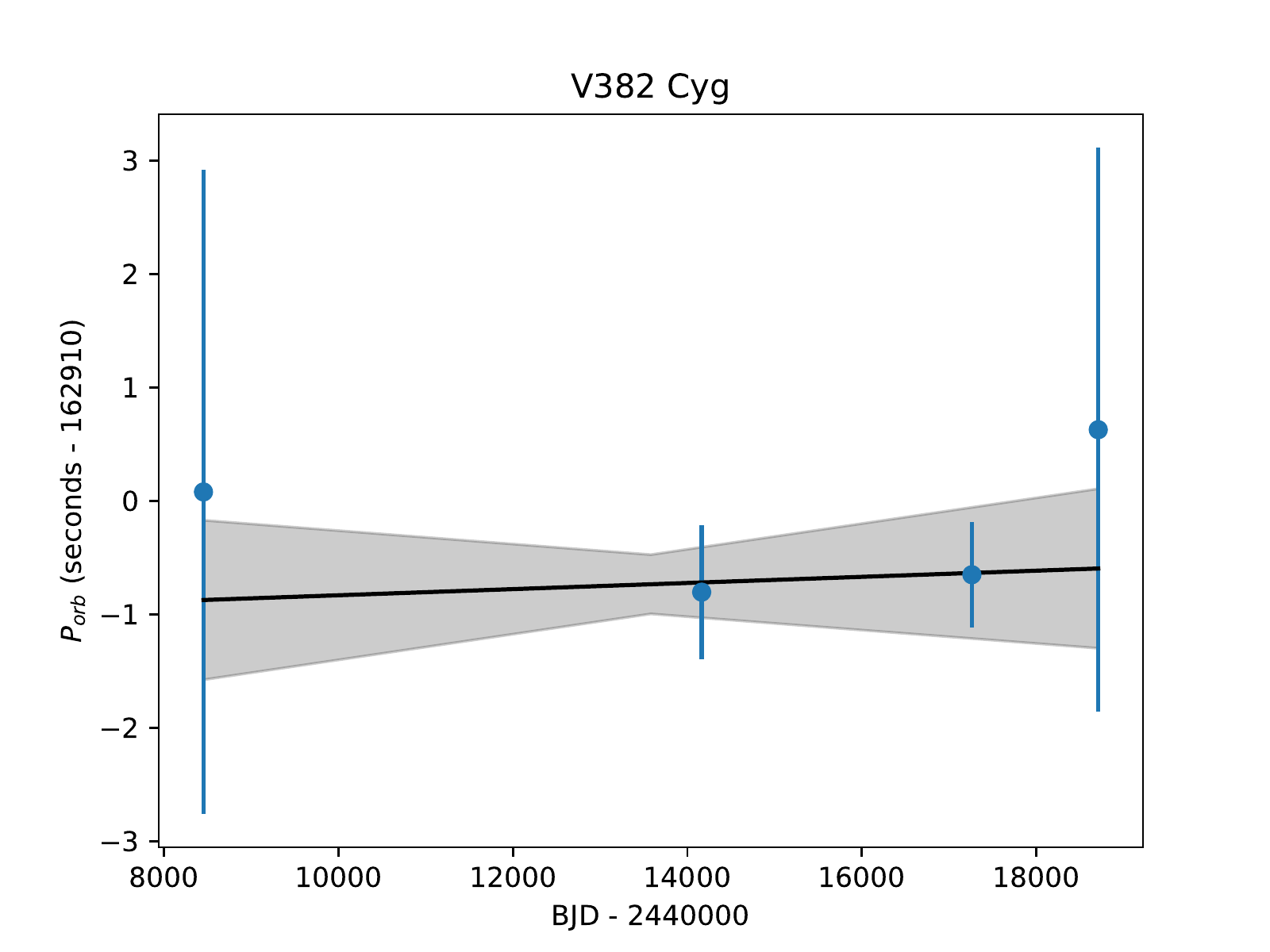}
    \caption{Measured period as a function of time for V382 Cyg.  The period is given in seconds and the period associated with the first observation, rounded to the nearest second, is subtracted.  A best fit line is plotted in black and its associated uncertainties are represented with the shaded region.}
    \label{fig:V382_Cyg_pdot}
    \end{figure}

Table \ref{table:periods} includes the central BJDs and orbital periods determined for each target from each independent data set together with their uncertainties.  Additionally, Table \ref{table:periods} also lists the $\dot{P}$ and corresponding errors as well as the $P/ |\dot{P}|$ for each object in our sample.  An example of a more graphical representation of our results can be found in Fig. \ref{fig:V382_Cyg_pdot}, which shows the measured periods for each of the data sets associated with V382 Cyg as well as the linear fit through these defining the $\dot{P}$. A similar figure for each of the other objects in our sample can be found in Appendix \ref{appendix1}.

In general, the measured periods were well constrained with small error bars (on order of 1 second or less) and agreed with one another within a few seconds for each object, with two notable exceptions.  In the case of LSS 3074, the errors on the period measurements were significantly larger than the rest of the sample by more than an order of magnitude, which in turn  led to a larger error on the derived $\dot{P}$. The other notable exception is SMC 108086, which showed small, but statistically significant downward trend over the time frame of the observations.

The measured $\dot{P}$ values for each object were all on the order of 0.1 seconds per year or less, with the exception of LSS 3074, which was about an order of magnitude higher.  That being said, SMC 108086 was the only object in our sample whose $\dot{P}$ measurement was not consistent with 0 within error.  Calculating $P/|\dot{P}|$, we find that most of our sample has period variation time scales of $\sim$1 Myr or larger, while LSS 3074 shows a variation timescale of closer to 0.3 Myr.  These values indicate that all objects in our sample are evolving on the nuclear timescale.  

Several previous works have computed the period changes for some of the objects in our sample using various methods, and in general, we find a very good agreement between our measurements and previous measurements.  In the case of V382 Cyg, there are a few independent period change measurements available in the literature \citep{Degirmenci1999,Qian2007, Yasarsoy2013}, and all indicate a period increase of between $\sim$ 0.03 and 0.04 seconds per year, which agrees with our measurement within error.  For VFTS 352 the period change was never directly measured, however \citet{Almeida2015} reports a peak to peak period difference of $\sim2$ seconds over a 12.5 year time frame, which corresponds to an upper limit of $|\dot{P}|\leq$ 0.16 seconds per year.  This value is in good agreement with the upper limit that we measure of 0.15 seconds per year.  Finally, TU Mus has one period change measurement in the literature from \citep{Qian2007}, who measured $\dot{P} = 0.035$ seconds per year, which agrees nicely with our measurement within error.

    \begin{table*}
    \caption{Measured periods for each of the data subsets for each system in the sample and the resulting period change and period stability.}
    \centering 
    
    \setlength{\extrarowheight}{6pt}
    \begin{tabular}{ccccccc}
    \hline\hline
    Object & Source & Central BJD & \multicolumn{2}{c}{Period} & $\dot{P}$ & $P/ |\dot{P}|$\\
     &  & (BJD - 2440000) & [d] & [s] & [s yr$^{-1}$] & [Myr] \\
    \hline
    \multirow{4}{*}{LSS 3074} & ANDICAM (V)  & 12015 & 2.1844 $\pm$ 0.0006 & 188730 $\pm$ 50 & \multirow{4}{*}{-0.7 $\pm$ 2.4} & \multirow{4}{*}{$0.27^{+\infty}_{-0.21}$}\\
                & ASAS (Ap3)  & 13483 & 2.185090 $\pm$ 0.000014 & 188791.8 $\pm$ 1.2 &  & \\
                & TESS (yr 1)  & 18610 & 2.1850 $\pm$ 0.0003 & 188780 $\pm$ 30 &  & \\
                & TESS (yr 3)  & 19347 & 2.1834 $\pm$ 0.0013 & 188650 $\pm$ 120 &  & \\
    \hline
    \multirow{4}{*}{MY Cam} & M\&V  & 14662 & 1.175476 $\pm$ 0.000004 & 101561.1 $\pm$ 0.3 & \multirow{4}{*}{-0.1 $\pm$ 0.6} & \multirow{4}{*}{$1.0^{+\infty}_{-0.8}$}\\
                & OMC (1/2)  & 14458 & 1.175427 $\pm$ 0.000006 & 101556.9 $\pm$ 0.5 &  & \\
                & OMC (2/2)  & 17770 & 1.175441 $\pm$ 0.000012 & 101558.1 $\pm$ 1.1 &  & \\
                & TESS (yr 2)  & 18828 & 1.17543 $\pm$ 0.00011 & 101557. $\pm$ 9. &  & \\
    \hline
    \multirow{7}{*}{SMC 108086} & OGLE II (I)  & 11250 & 0.883097 $\pm$ 0.000003 & 76299.6 $\pm$ 0.3 & \multirow{7}{*}{-0.11 $\pm$ 0.08} & \multirow{7}{*}{$0.7^{+2.1}_{-0.3}$}\\
                & OGLE III (I)  & 13529 & 0.883102 $\pm$ 0.000001 & 76300.04 $\pm$ 0.08 &  & \\
                & OGLE III (V)  & 14140 & 0.883084 $\pm$ 0.000003 & 76298.4 $\pm$ 0.2 &  & \\
                & OGLE IV (I)  & 16018 & 0.883089 $\pm$ 0.000003 & 76298.9 $\pm$ 0.2 &  & \\
                & OGLE IV (V)  & 16000 & 0.883103 $\pm$ 0.000014 & 76300.1 $\pm$ 1.2 &  & \\
                & TESS (yr 1)  & 18353 & 0.88306 $\pm$ 0.00002 & 76296. $\pm$ 2. &  & \\
                & TESS (yr 3)  & 19060 & 0.88301 $\pm$ 0.00002 & 76292.4 $\pm$ 1.8 &  & \\
    \hline
    \multirow{6}{*}{TU Mus} & Hipparcos  & 8411 & 1.387287 $\pm$ 0.000017 & 119861.6 $\pm$ 1.5 & \multirow{6}{*}{-0.005 $\pm$ 0.085} & \multirow{6}{*}{$25.4^{+\infty}_{-24.1}$}\\
                & OMC (1/2)  & 14443 & 1.387260 $\pm$ 0.000005 & 119859.3 $\pm$ 0.5 &  & \\
                & OMC (2/2)  & 17941 & 1.387290 $\pm$ 0.000005 & 119861.9 $\pm$ 0.5 &  & \\
                & ASAS (Ap3)  & 13527 & 1.387287 $\pm$ 0.000002 & 119861.6 $\pm$ 0.2 &  & \\
                & TESS (yr 1)  & 18596 & 1.387271 $\pm$ 0.000016 & 119860.3 $\pm$ 1.4 &  & \\
                & TESS (yr 3)  & 19333 & 1.387279 $\pm$ 0.000017 & 119860.9 $\pm$ 1.5 &  & \\
    \hline
    \multirow{4}{*}{V382 Cyg} & Hipparcos  & 8452 & 1.88553 $\pm$ 0.00003 & 162910. $\pm$ 3. & \multirow{4}{*}{0.01 $\pm$ 0.03} & \multirow{4}{*}{$16.5^{+\infty}_{-12.5}$}\\
                & OMC (1/2)  & 14163 & 1.885523 $\pm$ 0.000007 & 162909.2 $\pm$ 0.6 &  & \\
                & OMC (2/2)  & 17261 & 1.885525 $\pm$ 0.000005 & 162909.3 $\pm$ 0.5 &  & \\
                & TESS  & 18710 & 1.88554 $\pm$ 0.00003 & 162911. $\pm$ 2. &  & \\
    \hline
    \multirow{6}{*}{VFTS 352} & OGLE III (I)  & 13569 & 1.124167 $\pm$ 0.000001 & 97128.05 $\pm$ 0.10 & \multirow{6}{*}{-0.05 $\pm$ 0.10} & \multirow{6}{*}{$1.9^{+\infty}_{-1.2}$}\\
                & OGLE III (V)  & 13952 & 1.124154 $\pm$ 0.000005 & 97126.9 $\pm$ 0.5 &  & \\
                & OGLE IV (I)  & 15988 & 1.124151 $\pm$ 0.000001 & 97126.62 $\pm$ 0.12 &  & \\
                & OGLE IV (V)  & 15963 & 1.124162 $\pm$ 0.000003 & 97127.6 $\pm$ 0.3 &  & \\
                & TESS (yr 1)  & 18489 & 1.124195 $\pm$ 0.000005 & 97130.4 $\pm$ 0.4 &  & \\
                & TESS (yr 3)  & 19211 & 1.124172 $\pm$ 0.000006 & 97128.4 $\pm$ 0.5 &  & \\
    \hline
    \end{tabular}
    \label{table:periods}
    \end{table*}

\section{Discussion}\label{Discussion}

In order to assess our theoretical understanding of the past and future evolution of massive overcontact systems, we compare our observations with population synthesis results adapted from \cite{Menon2021}.   This population synthesis was originally computed from a grid of binary models corresponding to the metallicity of the Large Magellanic Cloud (LMC).  The original parameter space of the models spans an initial total mass of 20$-80$\,\msun, initial period of P$_\textrm{i} = 0.6-2$\,days and initial mass ratio of  q$_\textrm{i}=0.6-1$.  Given that this current work focuses on O+O overcontact systems, we only consider models that have  current primary and secondary masses $\geq14$\,\msun\, to compute the theoretical distribution of the observed parameters, namely $P_\textrm{orb}$, $q$ and $\dot{P}$. The reader is referred to \citet{Menon2021} for a more detailed description of the population synthesis computations.

    \begin{figure*}[ht]
    \centering
    \includegraphics[width=0.9\linewidth]{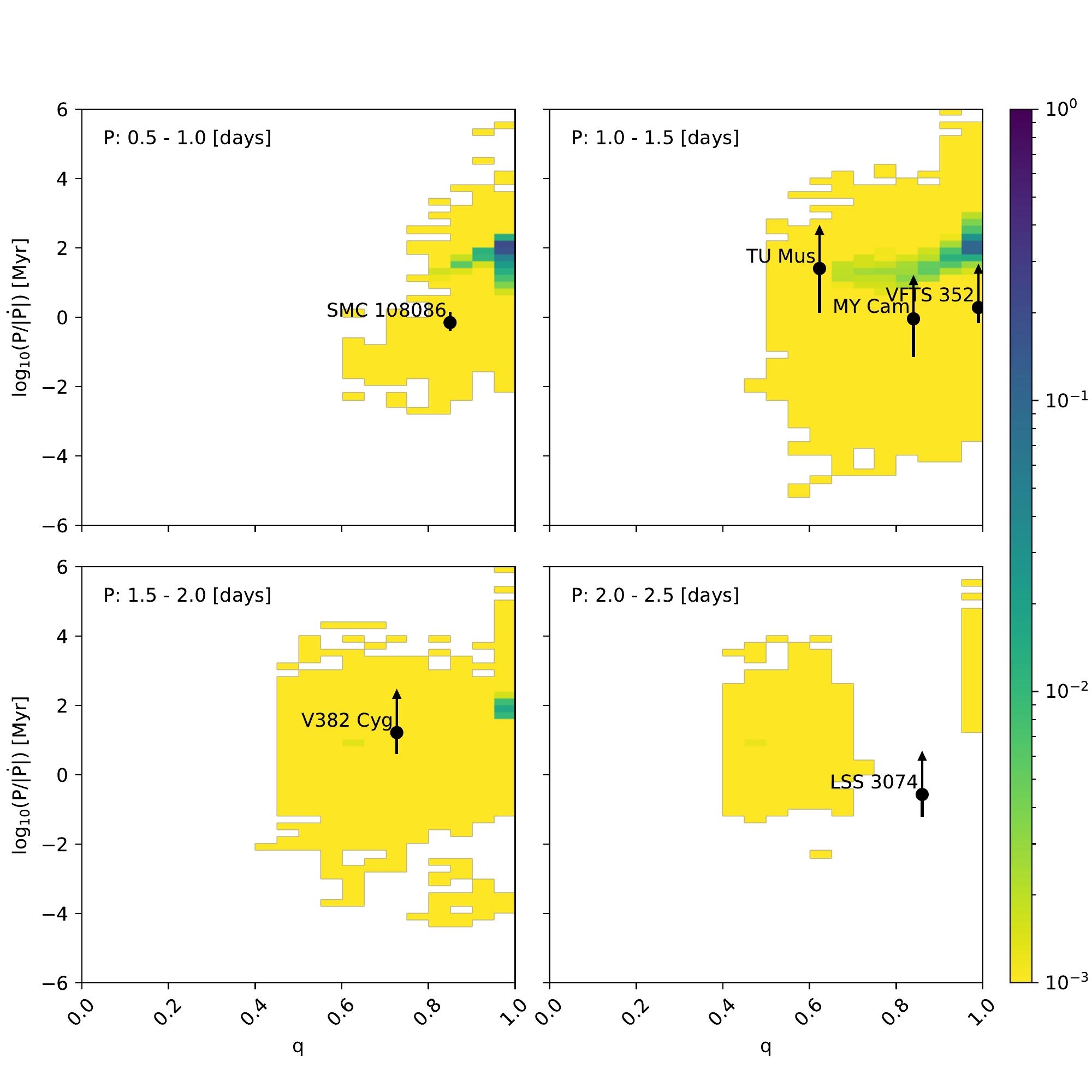}
    \caption{Normalized theoretical probability distribution of the $P / |\dot{P}|$ as a function of the mass ratio based on models from \citet{Menon2021}.  The background color represents the probability of finding a system with the given combination of parameters.  Lighter colors represent lower probabilities while darker colors represent higher probabilities.  Each of the four panels represents a different period bin, which is indicated in the upper left corner.  The locations of the observed overcontact systems are indicated with black dots and labeled.  Error bars are also plotted for each system and when applicable, arrows are used to indicate that the value does not have an upper limit.}
    \label{fig:q_vs_PPdot}
    \end{figure*}

In general, we find that the models are able to reproduce the orbital parameters of the observed systems.  However, based on Fig. \ref{fig:q_vs_PPdot}, it is clear that the observed systems do not follow the expected distribution as determined via population synthesis.  Almost all of the systems in the sample fall in low probability regions of the parameter space, indicating that these combinations of parameters are expected to be either very short-lived or rare.  

\begin{figure*}[ht]
    \centering
    \includegraphics[width=0.45\linewidth]{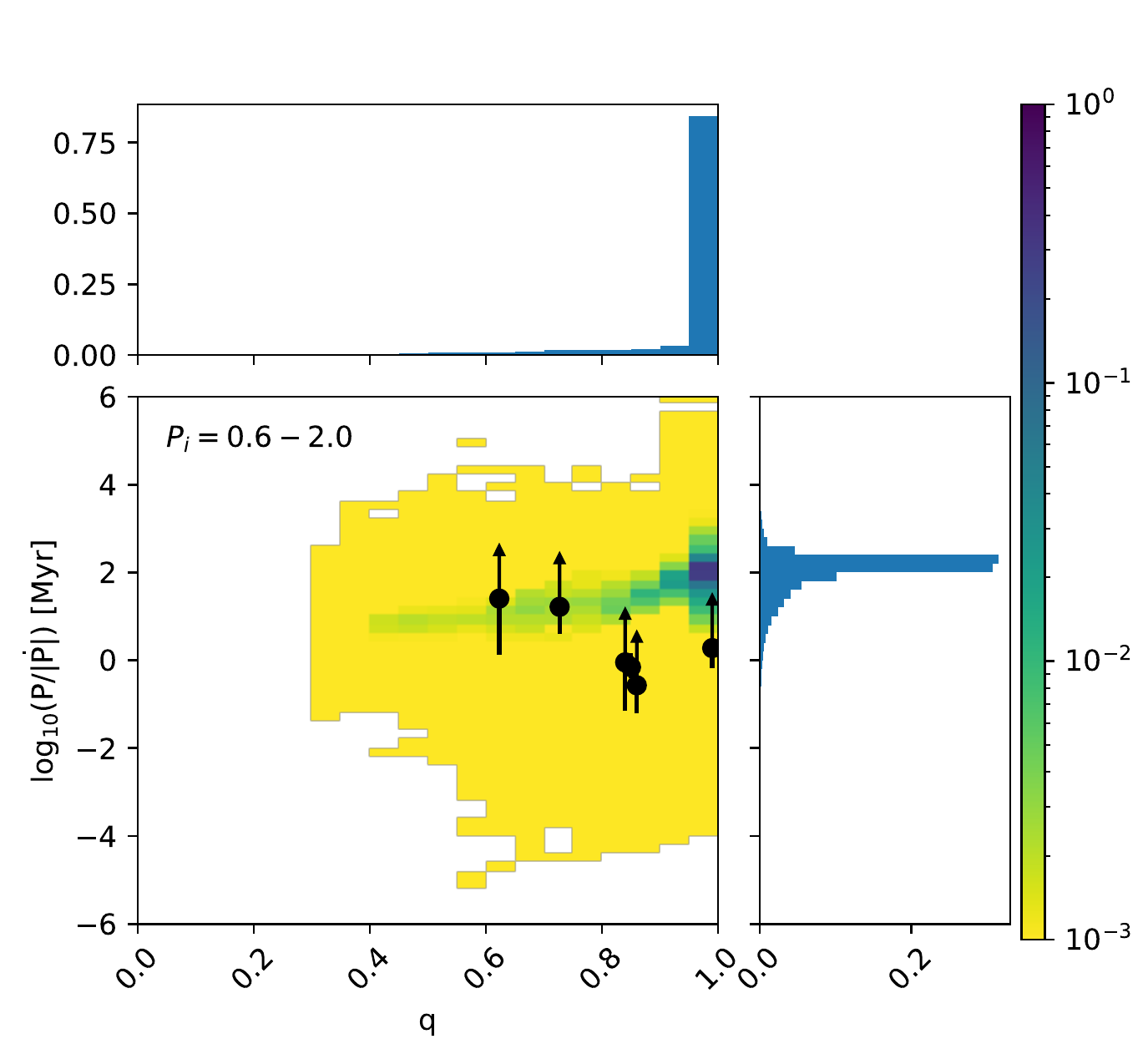}
    \includegraphics[width=0.45\linewidth]{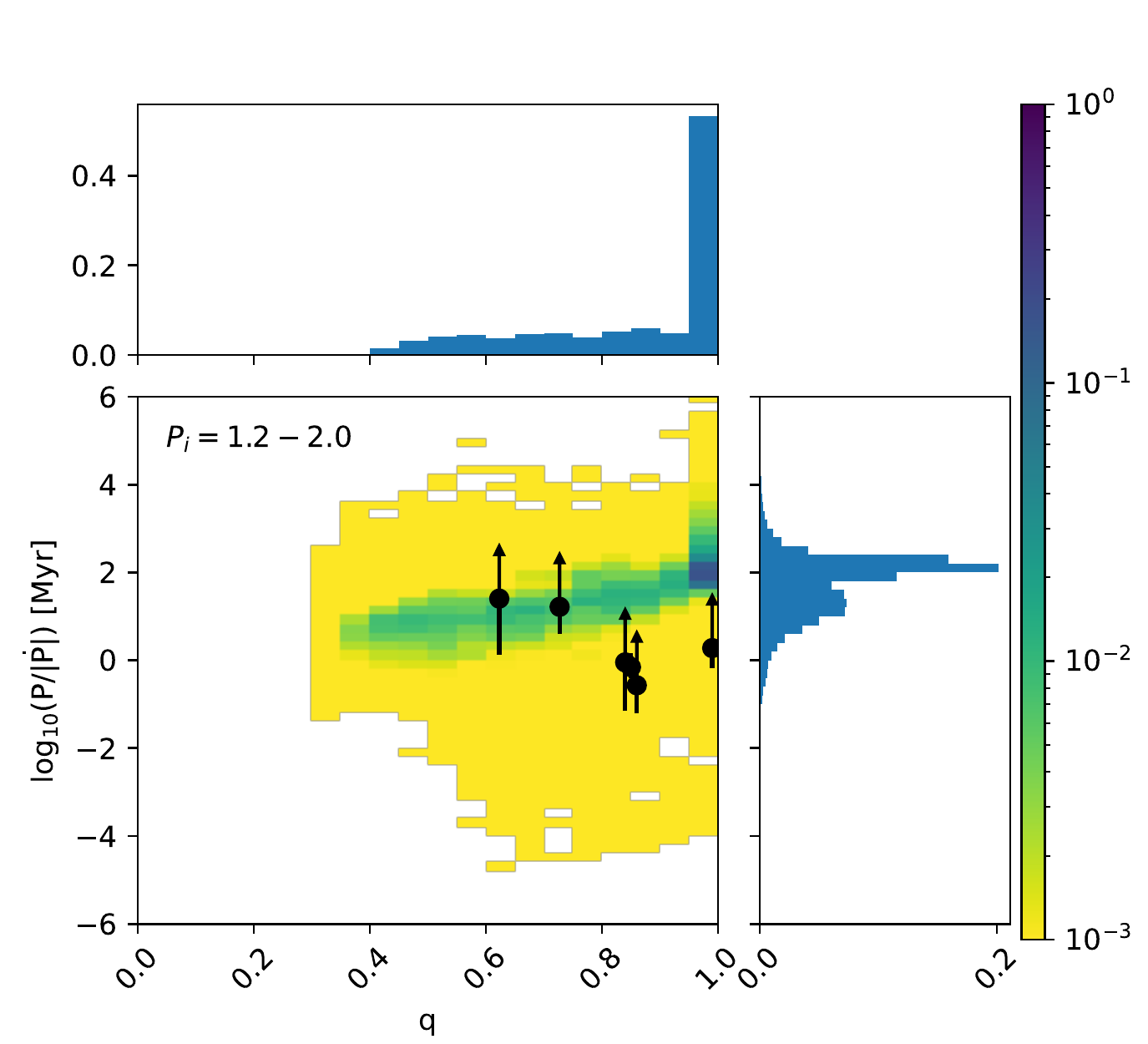}
    \caption{Same as Fig. \ref{fig:q_vs_PPdot} but for different initial period ranges:$P_i = 0.6-2.0$ and $P_i = 1.2-2.0$ for the left and right panels respectively. Additionally, 1D histograms corresponding to each of the axes are plotted.}
    \label{fig:P_i_comparison}
    \end{figure*}

A notable feature of the population synthesis results, which can be seen in Fig. \ref{fig:q_vs_PPdot} and the left panel of Fig \ref{fig:P_i_comparison}, is that the theoretical distribution for $P / |\dot{P}|$ peaks at around 100 Myrs, which is longer than the expected main sequence lifetime for stars in this mass range.  Given the fact that most of our measured $\dot{P}$ values are consistent with 0, this means that we are unable to rule out this possibility, however it is unlikely that these theoretical timescales are reliable.  The large  $P/|\dot{P}|$ values from the models are likely due to the way in which mass transfer is implemented during the contact phase in \texttt{MESA} \citep{Paxton2015,Marchant2016}. The mass transfer rate during the contact phase slows down to the order of 10$^{-7}$\,\msun/yr as soon as $q$  becomes close to 1, after which, the mass ratio asymptotically approaches $q = 1$ until the system finally merges \citep{Menon2021}. This causes the models to spend the majority of their main-sequence lives with mass ratios close to 1 as reflected in the theoretical distributions.  The observations on the other hand do not seem to support this, as the mass ratios are fairly well distributed between 0.6 and 1.  

Our observed mass ratio distribution is consistent with findings for lower mass contact systems as well, where low mass ratio contact systems are common \citep[see e.g., ][ and references therein]{Yang2015, Qian2020}.  In studies of low mass convective core contact systems, observations have shown that the period stabilities are comparable to the values that we find here. Further, several systems have period changes suggesting that they are evolving towards a lower mass ratio rather than to 1 \citep{Yang2015}.

Among the O+O models, we find that the main source of the q = 1 contact binaries are models with initial periods $P_\textrm{i} \leq 1.2$\,days. If we only consider models with initial periods larger than 1.2\,days, while the peak of the distribution still lies at q = 1, the  distribution flattens considerably over the $q$ dimension, and we begin to see a clear correlation between the mass ratio and the period stability (see Fig. \ref{fig:P_i_comparison}). Interestingly, however, this correlation does not appear to be present in the observed distribution, suggesting that these systems may not equalize on the timescales that the models predict. That being said, these findings may suggest that the observed overcontact binaries are originating from systems with longer initial periods \citep[in line with findings from ][]{Ramirez-Tannus2017, Ramirez-Tannus2021, Sana2017},  however a more dedicated theoretical investigation is needed to confirm this hypothesis.


While there appears to be a definite discrepancy between the observed population and the one predicted from population synthesis, there are several factors that should be considered before drawing conclusions:  

First, it should be noted that the models from \citet{Menon2021} are calculated assuming LMC metallicity, while most of our sample is Galactic.  This difference in metallicity could affect the periods and period stabilities as massive stars at higher metallicities tend to have slightly larger radii and stronger winds at the same evolutionary stage, which may lead to shorter overall period stabilities. 

Additionally, the population synthesis results assume systems that have an initial period of two days or less and assume that all mass transfer is conservative.  Given that higher initial periods seem to allow for a more even q distribution, including initial periods of greater than two days in the population synthesis could allow for a better agreement between the population synthesis results and the observations.  

Furthermore, the population synthesis results assume the systems have an initial mass ratio of greater than 0.6.  As discussed in \citet{Menon2021}, the likelihood of a system coming into contact, as well as the duration of the contact phase are strongly correlated with the initial mass ratio, implying that these systems would represent a small minority of the currently observable contact systems. That being said, the inclusion of systems with lower initial mass ratios may allow for a marginally better agreement between the population synthesis results and the observations.

An additional factor to consider is that the binary models do not include energy transfer.  Considering the mass-radius relationship of single stars, as well as the strict relationship on their radii when a system is in contact, stable overcontact systems with a mass ratio away from unity would not be expected to exist theoretically \citep{Kuiper1941}. However, as energy transfer is expected to occur in overcontact layers, the mass-radius relationship becomes dependent on the mass ratio and separation of the system, potentially allowing for stable solutions to exist \citep[see e.g., ][]{shu1976}. A detailed analysis on the impact of energy transfer on populations of massive overcontact binaries has, however, not been done yet. 

Finally, the implementation of the contact scheme itself in \texttt{MESA}, leads to the binary model spending an inordinately large amount of its contact lifetime close to a mass ratio of q = 1. This may indicate the requirement to improve the current contact scheme used in our models.  While each of these assumptions will surely affect the final distribution, it is unlikely that the changes would be significant enough to rectify the discrepancy between the observations and the theoretical predictions.  This could however account for the mass ratio gap that is seen in the bottom right panel of Fig. \ref{fig:q_vs_PPdot}, and could perhaps allow the models to reproduce the location of LSS 3074.

One additional point to consider involves the comparison of the observed period stability with the theoretical values.  As discussed in \citet{King2021}, the measured $P / |\dot{P}|$ may be misleading on small time scales as changes in period can be caused by variations on the flow or temporary digressions from synchronicity.  Over the long term, these fluctuations would average out, allowing a more robust comparison with theoretical models.  It should be noted, however that \citet{King2021} and studies like it \citep[see e.g.][]{Pringle1975} focus on ultraluminous X-ray (ULX) sources, where the primary stars are overflowing through L1, transferring mass to their companions.  It is unlikely that overcontact systems would suffer from the same level of period variations as ULX sources given that overcontact systems are expected to be in hydrostatic equilibrium and rotating synchronously.  Nevertheless, comparing the $P / |\dot{P}|$ of the complete sample of O+O overcontact systems as a whole instead of individual sources allows us to circumvent this potential issue.

\section{Conclusions}\label{Conclusions}

We have performed a period stability study of known O+O type overcontact systems.  Using archival photometric data and the software package PERIOD04, we calculated the periods of the systems over a time span of tens of years.  For each system in our sample, we determined the rate at which the period is changing via a linear regression through the period measurements of each data subset.  We find that all systems in our sample show period changes consistent with 0 with the exception of SMC 108086, which shows a slight but non-negligible negative period change.   These results indicate that all of the systems in our sample have periods that are stable on the nuclear timescale.  Furthermore, we find no correlation between the mass ratio and the period stability, implying that these systems will continue to evolve as unequal mass overcontact binaries.

Comparing our results with population synthesis simulations, we find discrepancies between the predicted and observed distributions.  While the population synthesis simulations predict that the overwhelming majority of overcontact systems should be found in equal mass systems, the mass ratios of the observed systems are fairly evenly distributed between $q = $ 0.6 and 1. This discrepancy is marginally lessened by removing the shortest period systems in the population synthesis simulations, suggesting that the observed population of overcontact systems may have originated from binaries with longer initial periods.  A more in depth theoretical investigation is needed to confirm this, however.  That being said, without a larger sample size, it is difficult to draw strong conclusions, highlighting the need for a dedicated effort to search for and characterize currently undiscovered massive overcontact systems.

\begin{acknowledgements}
      This paper includes data collected by the TESS mission, which are publicly available from the Mikulski Archive for Space Telescopes (MAST). Funding for the TESS mission is provided by NASA’s Science Mission directorate.
      This research made use of Lightkurve, a Python package for Kepler and TESS data analysis \citep{LightkurveCollaboration2018}.
      AM would like to thank the Alexander von Humboldt foundation for supporting this project.
      L.M. thanks the European Space Agency (ESA) and the Belgian Federal Science Policy Office (BELSPO) for their support in the framework of the PRODEX Programme.
      PM acknowledges support from the FWO junior postdoctoral fellowship No. 12ZY520N
\end{acknowledgements}

\bibliographystyle{aa}
\bibliography{mybib}

\appendix

\section{Period plots} \label{appendix1}

    \begin{figure}[ht]
    \centering
    \includegraphics[width=\linewidth]{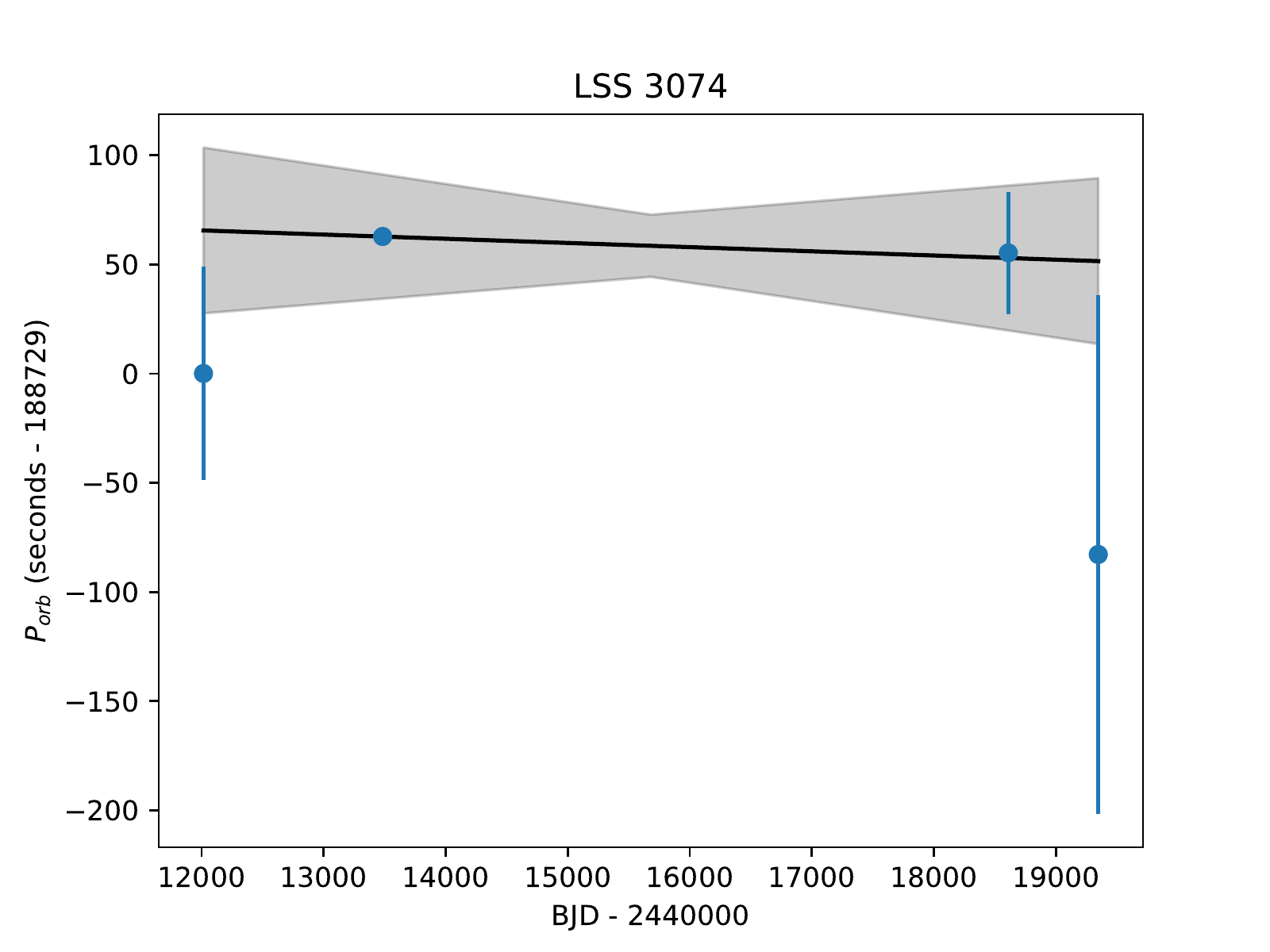}
    \caption{Same as Fig. \ref{fig:V382_Cyg_pdot} but for LSS 3074.}
    \label{fig:LSS_3074_pdot}
    \end{figure}
    
    \begin{figure}[ht]
    \centering
    \includegraphics[width=\linewidth]{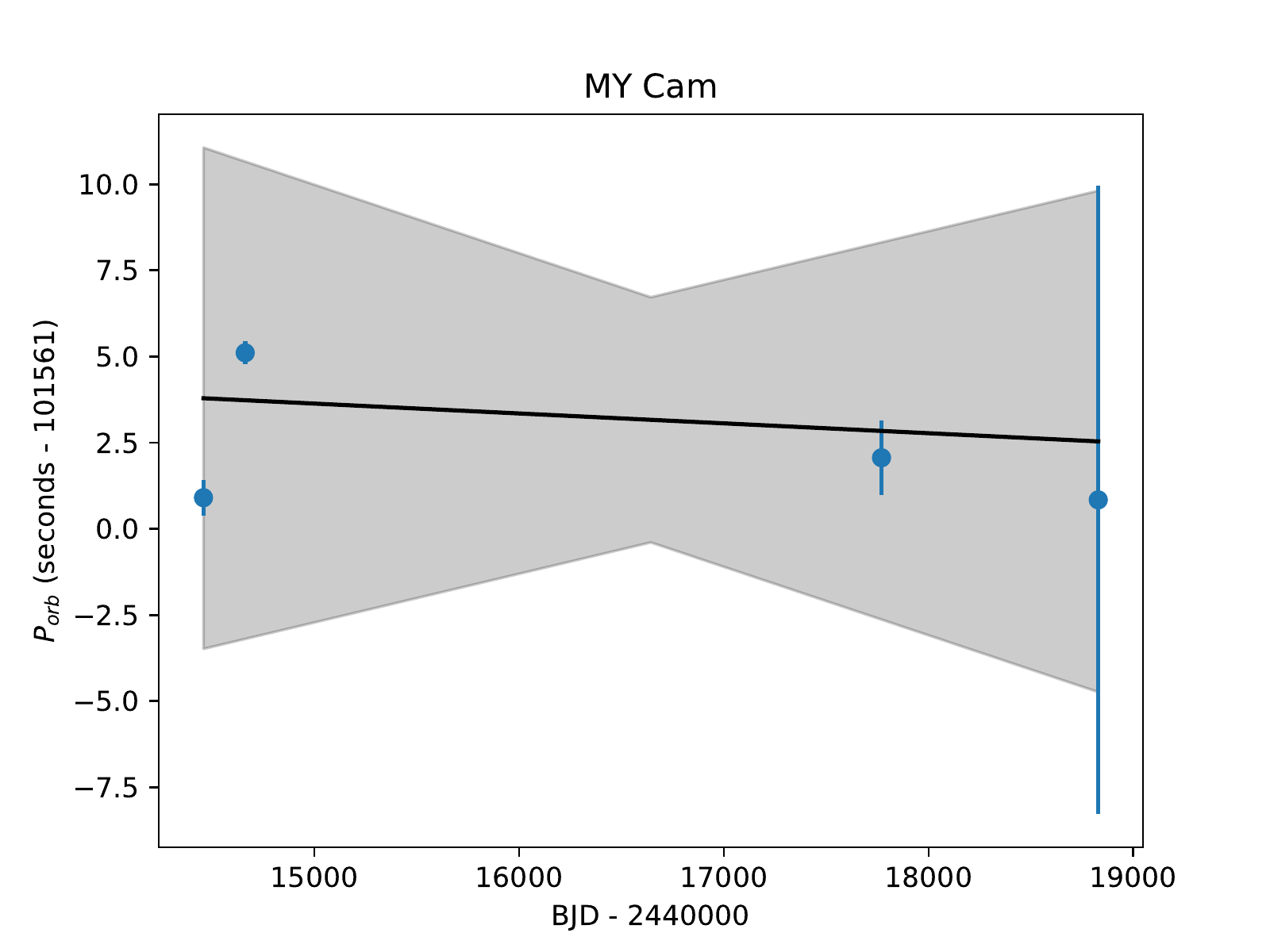}
    \caption{Same as Fig. \ref{fig:V382_Cyg_pdot} but for MY Cam.}
    \label{fig:MY_Cam4_pdot}
    \end{figure}
    
    \begin{figure}[ht]
    \centering
    \includegraphics[width=\linewidth]{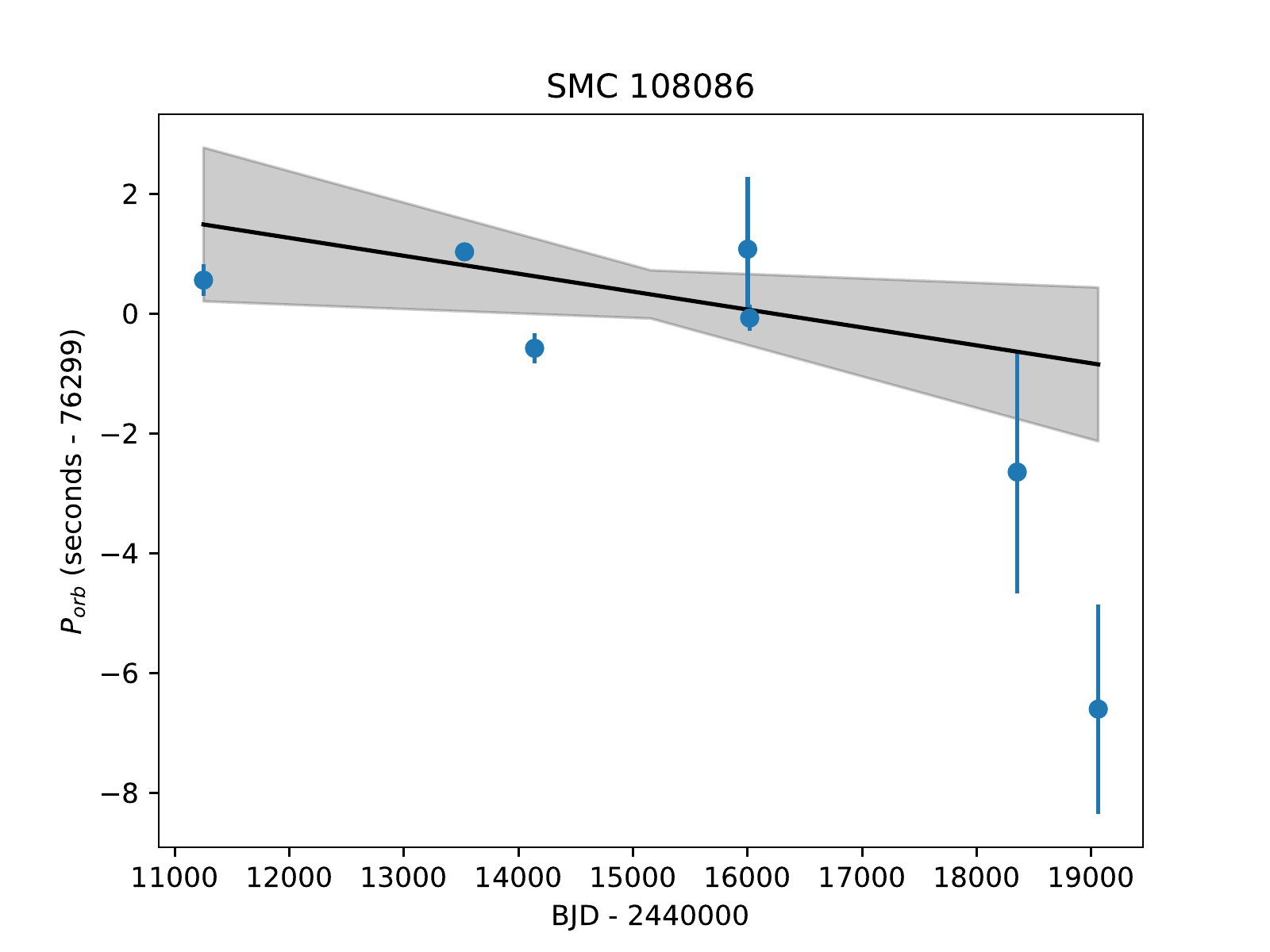}
    \caption{Same as Fig. \ref{fig:V382_Cyg_pdot} but for SMC 108086.}
    \label{fig:SMC_108086_pdot}
    \end{figure}
    
    \begin{figure}[ht]
    \centering
    \includegraphics[width=\linewidth]{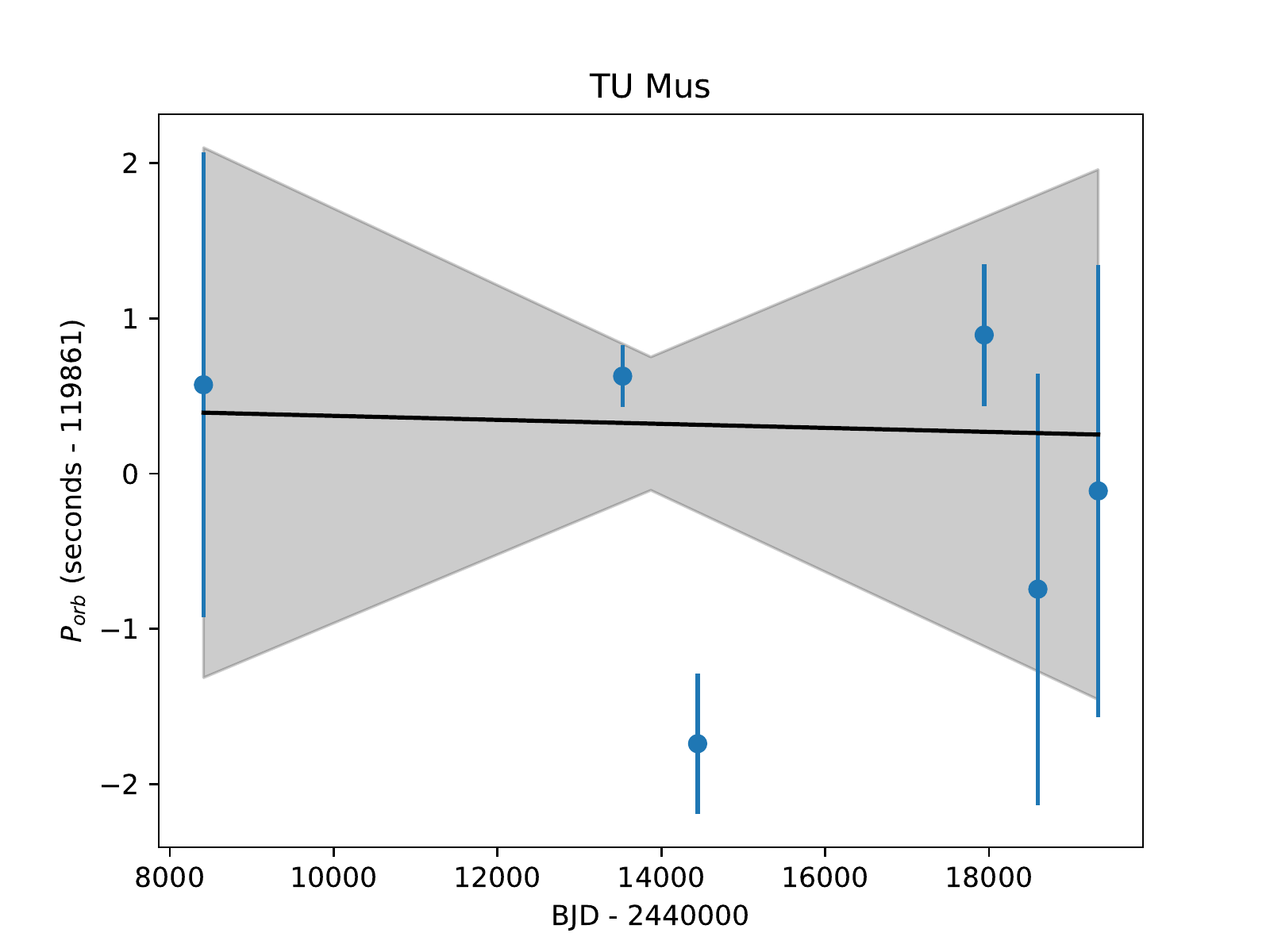}
    \caption{Same as Fig. \ref{fig:V382_Cyg_pdot} but for V382 Cyg.}
    \label{fig:TU_Mus_pdot}
    \end{figure}

    \begin{figure}[ht]
    \centering
    \includegraphics[width=\linewidth]{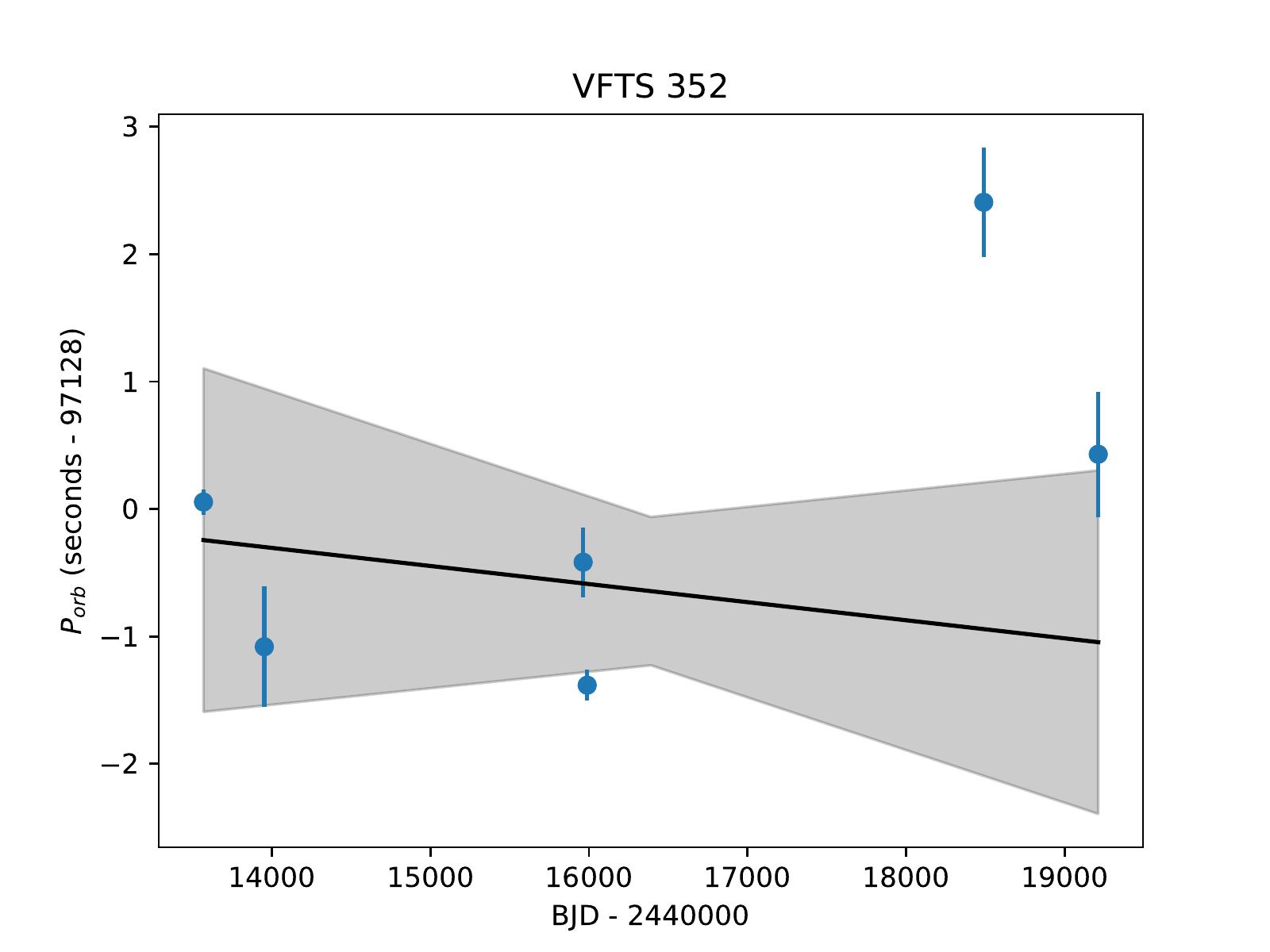}
    \caption{Same as Fig. \ref{fig:V382_Cyg_pdot} but for VFTS 352.}
    \label{fig:VFTS_352_pdot}
    \end{figure}

\end{document}